\documentstyle[12pt,epsf]{article}
\begin{document}

\begin{titlepage}
\rightline{\vbox{\halign{&#\hfil\cr
&UQAM-PHE-98/07\cr
&\today\cr}}}
\vspace{0.5in}
\begin{center}

\Large\bf  
Higgs-mediated FCNC in Supersymmetic Models with Large $\tan\beta$.
\\
\medskip
\vskip0.5in

\normalsize {{\bf C. Hamzaoui}\footnote{hamzaoui@mercure.phy.uqam.ca}, 
{\bf M. Pospelov}}\footnote{pospelov@mercure.phy.uqam.ca}
and {\bf M. Toharia}\footnote{toharia@mercure.phy.uqam.ca}
\smallskip
\medskip

{ \sl 
D\'{e}partement de Physique, Universit\'{e} du Qu\'{e}bec {\`a} 
Montr\'{e}al\\ 
C.P. 8888, Succ. Centre-Ville, Montr\'{e}al, Qu\'{e}bec, 
Canada, H3C 3P8} 
\smallskip
\end{center}
\vskip1.0in

\noindent
{\large\bf Abstract}
\smallskip\newline
In the supersymmetric models with nontrivial flavour 
structure in the soft-breaking sector the exchange of neutral Higgses mediates $\Delta F=2$ transitions. This mechanism is studied for $\Delta S, \Delta B=2$ processes and for a generic form of the soft-breaking terms. We find that Higgs-mediated FCNC amplitudes increase very rapidly with $\tan\beta$ and can exceed $SUSY$ box contribution by up to two orders of magnitude when $\tan\beta\sim m_t/m_b$.  
\end{titlepage}

\baselineskip=20pt

\newpage \pagenumbering{arabic}

\section{Introduction}

\smallskip

The supersymmetric version of Standard Model around the electroweak scale
has been a subject of very elaborate theoretical investigations during the past
two decades \cite{SUSY}. In this article we study Flavour Changing Neutral Current (FCNC)
processes associated with Higgs exchange as the consequence of the
supersymmetric threshold corrections to the Yukawa interaction in a generic $%
SUSY$ model.

\smallskip The interest to the supersymmetric models with large $\tan \beta $
comes from the simple idea to attribute the huge difference between the observed masses $m_{b}$ and $m_{t}$ to the possible hierarchy in the Higgs vevs, $%
v_{u}/v_{d}\sim m_{t}/m_{b}.$ This means that the Yukawa couplings of the
top and bottom quarks can be comparable or even equal, $y_{b}=y_{t}.$
Supersymmetric models with large $\tan \beta $ predict significant
contributions to the low energy observables \cite{LTB} such as the anomalous magnetic
moment of the muon, $b\rightarrow s\gamma $ branching ratio and other
processes with linear dependence of $\tan \beta .$ The radiative corrections 
$\Delta m_{d,s,b}$ to the masses of the Down-type quarks are also
significant and may account for up to 100\% of the observable masses \cite{TC}.

As to the potential influence of the soft-breaking sector on the mass
splitting in neutral $K$ and $B$ mesons, it is commonly attributed to the
box diagram with the superpartners inside and nontrivial flavour structure of
the scalar quark mass matrices \cite{DGH,FCNC}. The box diagrams provide very important but
not the only source of the non-SM contributions to the $\Delta F=2$
amplitudes. Another class of contributions is related with the
supersymmetric corrections to the Higgs potential. These corrections change
the familiar Yukawa interaction generated by the superpotential making it
similar to generic two-Higgs doublet type of interaction with the presence
of $H_{u}^{*}$ and $H_{d}^{*}$ fields. It is clear that this two-doublet
model will possess FCNC mediated by neutral Higgses {\em if }the squark
sector has nontrivial flavour structure.

How large these Higgs-mediated FCNC amplitudes could be? This can be
understood qualitatively in a SUSY theory with a generic form
of squark mass matrices. In this case a simple estimate of $\Delta F=2$ amplitudes
in the Down quark sector induced by  box diagrams gives $(\alpha
_{s})^{2}m_{sq}^{-2}$ times the flavour changing piece in the soft-breaking sector $\Delta m^2_{ij}/m_{sq}^2$ which we take here to be of the order 1. Here $m_{sq}$ is a typical soft-breaking mass. On the
other hand, the tree level amplitudes mediated by Higgses behave as $%
y_{b}^{2}m_{higgs}^{-2}$ times some power of mixing angles between the quark
mass basis dictated by the superpotential and the physical basis which
includes $v_{u}$ corrections. These angles can be estimated as $\Delta
m_{b}/m_{b}\sim (\alpha _{s}/3\pi )\mu m_{sq}^{-1}(v_{u}/v_{d})$ and FCNC
amplitudes mediated by Higgses are given by the expression: 
\begin{equation}
FCNC_{Higgs}\sim \frac{y_{b}^{2}}{m_{higgs}^{2}}\left[ \frac{\alpha _{s}\tan
\beta }{3\pi }\frac{\mu }{m_{sq}}\right] ^{n}
  \label{estim}
\end{equation}
with $n=2$ for $\Delta B=2$ and $n=4$ for $\Delta S=2$
processes. To compare it with the box-induced FCNC amplitude we take $\mu
\sim m_{higgs}\sim m_{sq}$ and $y_{b}\sim \tan \beta /60.$ As a result, we
obtain the following relation between two FCNC mechanisms: 
\begin{equation}
\frac{FCNC_{Higgs}}{FCNC_{box}}\sim \left[ \frac{\tan \beta }{60\alpha _{s}}%
\right] ^{2}\left[ \frac{\alpha _{s}\tan \beta }{3\pi }\right] ^{n}.
\label{comp}
\end{equation}

If $\tan \beta $ is sufficiently large, $FCNC_{Higgs}$ provides the dominant
contribution to $\Delta F=2$ processes in the Down sector and the estimates of
the critical values for $\tan \beta $ are: 
\begin{eqnarray}
\left( \tan \beta \right) _{cr} &\sim &\sqrt{180\pi }\sim 25\mbox{ and }%
\frac{FCNC_{Higgs}}{FCNC_{box}}\sim \left[ \frac{\tan \beta }{25}\right] ^{4}%
\mbox{for }\Delta B=2, \\
\left( \tan \beta \right) _{cr} &\sim &\left( 540\pi ^{2}\alpha
_{s}^{-1}\right) ^{1/3}\sim 35\mbox{ and }\frac{FCNC_{Higgs}}{FCNC_{box}}%
\sim \left[ \frac{\tan \beta }{35}\right] ^{6}\mbox{\ for }\Delta S=2.
\end{eqnarray}

Although these critical values appear to be quite large, the strong growth of $%
FCNC_{Higgs}$ with $\tan \beta $ makes the Higgs-mediated amplitudes one to two
orders of magnitude larger than the $SUSY$ box contribution when $\tan \beta
\sim 60$.\thinspace The growth with $\tan \beta $ stabilizes at some point
and for $\tan \beta \rightarrow \infty $ (purely radiative mechanism for $%
M_{d}$) reaches a certain saturation limit.

These qualitative arguments make us believe that the problem of FCNC in the
large $\tan \beta $ regime deserves special consideration. In the next
section we will construct $\Delta F=2$ FCNC amplitudes mediated by Higgs exchange and
put the limits on the flavour changing terms in the scalar quark mass
matrices. This will be followed by the constraints on the minimal
supergravity model, supersymmetric $SO(10)$ GUT and supersymmetric
left-right models imposed by the Higgs-mediated FCNC.

\section{Higgs-mediated FCNC and the limits on the soft-breaking sector}

The superpotential of the minimal supersymmetric standard model (MSSM), 
\begin{equation}
W=\epsilon _{ij}[-Q^{i}H_{2}^{j}{\bf Y}_{u}^{(0)}U+Q^{i}H_{1}^{j}{\bf Y}%
_{d}^{(0)}D+L^{i}H_{1}^{j}{\bf Y}_{e}^{(0)}E+\mu H_{1}^{i}H_{2}^{j}],
\label{w}
\end{equation}
contains the same number of free dimensionless parameters as the Yukawa
sector of the standard model. All the interactions of the Higgs particles
with fermions conserve flavour. For simplicity, we choose the basis where $%
{\bf Y}_{d}^{(0)}$ is taken in the diagonal form, ${\bf Y}%
_{d}^{(0)}=diag(y_{d}^{(0)};y_{s}^{(0)};y_{b}^{(0)})$.

The soft-breaking sector can influence flavour physics. Among different
scalar masses, the soft-breaking sector has the squark mass terms 
\begin{equation}
\tilde{U}^{\dagger }{\bf M}_{U}^{2}\tilde{U}+\tilde{D}^{\dagger }{\bf M}%
_{D}^{2}\tilde{D}+\tilde{Q}^{\dagger }{\bf M}_{Q}^{2}\tilde{Q};
\end{equation}
and the trilinear terms 
\begin{equation}
\epsilon _{ij}\left( -\tilde{Q}^{i}H_{2}^{j}{\bf A}_{u}\tilde{U}+\tilde{Q}%
^{i}H_{1}^{j}{\bf A}_{d}\tilde{D}\right) ~+H.c.;  \label{eq:A}
\end{equation}
as the possible sources of flavour transitions.

Below the supersymmetric threshold the Yukawa interaction of the quarks with
the Higgs fields has the generic form of the two-Higgs doublet model:

\begin{eqnarray}
-{\cal L}_{Y} &=&\overline{U}_{R}Q_{L}H_{u}({\bf Y}_{u}^{(0)}+{\bf Y}%
_{u}^{(1)})+\overline{U}_{R}Q_{L}H_{d}^{*}{\cal Y}_{u}+\overline{D}%
_{R}Q_{L}H_{d}({\bf Y}_{d}^{(0)}+{\bf Y}_{d}^{(1)})  \label{2h} \\
&&+\overline{D}_{R}Q_{L}H_{u}^{*}{\cal Y}_{d}+h.c.,  \nonumber
\end{eqnarray}
and $SU(2)$ indices are suppressed here.

The diagrams generating ${\cal Y}_{d}$ are shown in Fig. 1. The radiative corrections
are of the order of the tree level contribution  if $\tan \beta $ is large and $\mu \sim m_{sq}\sim
m_{\lambda }$ \cite{TC}. As a result, the equality of Yukawa couplings for top and
bottom quarks, $y_{b}^{(0)}=y_{t}^{(0)}$, does not fix $\tan \beta $, making
it rather complicated function of $\mu ,\;m_{sq},$\ $m_{\lambda }$ and the
relative phase between the tree-level contribution and the radiative
correction piece \cite{HP}. It is clear that the same diagrams generate the corrections to the mixing angles if flavour can be changed on the squark line. 

To find out wether Higgs particles mediate FCNC, we need to know the quark mass
matrices, 
\begin{eqnarray}
\sqrt{2}M_{u} &=&({\bf Y}_{u}^{(0)}+{\bf Y}_{u}^{(1)})v_{u}+{\cal Y}_{u}v_{d}
\nonumber \\
\sqrt{2}M_{d} &=&({\bf Y}_{d}^{(0)}+{\bf Y}_{d}^{(1)})v_{d}+{\cal Y}%
_{d}v_{u}\simeq {\bf Y}_{d}^{(0)}v_{d}+{\cal Y}_{d}v,  \label{mful}
\end{eqnarray}
and the decomposition of the neutral Higgs fields into the physical
eigenstates \cite{HHG}: 
\begin{eqnarray}
H_{u}^{0} &=&v_{u}+H^{0}\sin \alpha +h^{0}\cos \alpha +iA^{0}\cos \beta
+iG^{0}\sin \beta \\
H_{d}^{0} &=&v_{d}+H^{0}\cos \alpha -h^{0}\sin \alpha +iA^{0}\sin \beta
-iG^{0}\cos \beta .  \nonumber
\end{eqnarray}
In the formula for $M_d$ (\ref{mful}) we have neglected the radiative correction piece proportional to $v_d$ and put $v_u$ equal to the SM Higgs vev $v$. 

The angles $\alpha $ and $\beta $ here are connected as follows: 
\begin{equation}
\sin 2\alpha =-\sin 2\beta \frac{m_{H}^{2}+m_{h}^{2}}{m_{H}^{2}-m_{h}^{2}}%
;\;\;\cos 2\alpha =-\cos 2\beta \frac{m_{A}^{2}-m_{Z}^{2}}{%
m_{H}^{2}-m_{h}^{2}}.
\end{equation}

The presence or absence of Higgs-mediated FCNC in the Down quark sector
depends on the particular forms of the matrices $M_{d},{\bf Y}_{d}^{(0)},{\bf Y}%
_{d}^{(1)}$ and ${\cal Y}_{d}$. But even before analyzing possible forms of
the mass matrices and Yukawa couplings, we are able to present four-fermion
operators which can produce leading effects. These are operators which have
the square of the initial Yukawa coupling of the $b^{(0)}$-quark, the
eigenstate of the Yukawa coupling ${\bf Y}_{d}^{(0)}$ from the superpotential:

\begin{eqnarray}
-{\cal L}_{4f} &=&\overline{b}_{L}^{(0)}b_{R}^{(0)}\overline{b}%
_{R}^{(0)}b_{L}^{(0)}|y_{b}^{(0)}|^{2}\left( \frac{\cos ^{2}\alpha }{m_{H}^{2}}+\frac{\sin ^{2}\beta }{m_{A}^{2}}+\frac{\sin ^{2}\alpha }{m_{h}^{2}}%
\right)  \nonumber \\
&&+\frac{1}{2}\overline{b}_{L}^{(0)}b_{R}^{(0)}\overline{b}%
_{L}^{(0)}b_{R}^{(0)}|y_{b}^{(0)}|^{2}\left( \frac{\cos ^{2}\alpha }{m_{H}^{2}}-\frac{\sin ^{2}\beta }{m_{A}^{2}}+\frac{\sin ^{2}\alpha }{m_{h}^{2}}%
\right) \label{b0} \\
&&+\frac{1}{2}\overline{b}_{R}^{(0)}b_{L}^{(0)}\overline{b}%
_{R}^{(0)}b_{L}^{(0)}|y_{b}^{(0)}|^{2}\left( \frac{\cos ^{2}\alpha }{m_{H}^{2}}-\frac{\sin ^{2}\beta }{m_{A}^{2}}+\frac{\sin ^{2}\alpha }{m_{h}^{2}}%
\right) \nonumber
\end{eqnarray}

In the large $\tan \beta $ regime, $\sin ^{2}\alpha \simeq $ $\tan ^{-2}\beta
; $ $m_{H}^{2}\simeq m_{A}^{2}$ and therefore the relevant part of this
interaction is:

\begin{eqnarray}
-{\cal L}_{4f} &=&\overline{b}_{L}^{(0)}b_{R}^{(0)}\overline{b}%
_{R}^{(0)}b_{L}^{(0)}|y_{b}^{(0)}|^{2}\frac{2}{m_{A}^{2}} \\
&&+\frac{1}{2}\left( \overline{b}_{L}^{(0)}b_{R}^{(0)}\overline{b}%
_{L}^{(0)}b_{R}^{(0)}+\overline{b}_{R}^{(0)}b_{L}^{(0)}\overline{b}%
_{R}^{(0)}b_{L}^{(0)}\right) |y_{b}^{(0)}|^{2}\frac{1}{m_{h}^{2}\tan
^{2}\beta }.  \nonumber
\label{b0b0}
\end{eqnarray}
Here we have assumed that $m_H,~m_A\gg m_Z,~m_h$ and neglected further $1/\tan ^{2}\beta $ and $m_h^2/m_H^2$ suppressed terms.

The diagonalization of the Down-quark mass matrix (\ref{mful}) by the
bi-unitary transformation 
\begin{equation}
\sqrt{2}M_{d}^{diag}=U^{\dagger }({\bf Y}_{d}^{(0)}v_{d}+{\cal Y}_{d}v)V,
\label{Md}
\end{equation}
sets the physical (mass) basis for the Down quarks. Matrix $V$ rotates
left-handed fields and, therefore, renormalizes initial Kobayashi-Maskawa
matrix $V_{KM}^{(0)}$ provided by the non-commutativity of ${\bf Y}_{u}^{(0)}
$ and ${\bf Y}_{d}^{(0)}$ in the superpotential, 
\begin{equation}
V_{KM}=V_{KM}^{(0)}V.
\end{equation}

This by itself can be considered as a certain constraint on $V$ if we assume
that no fine tuning occurs in the product of $V_{KM}^{(0)}$ and $V$ so that $%
|V_{31}|<|(V_{KM})_{td}|$; $|V_{32}|<|(V_{KM})_{ts}|,$ etc. As to the right-handed rotation matrix $U$, it appears absolutely arbitrary and its mixing angles can be limited only through possible FCNC contributions associated with them. The same bi-unitary transformation
brings the off-diagonal entries into the four-fermion interaction (\ref{b0b0}):

\begin{eqnarray}
-{\cal L}_{4f} &=&\overline{D}_{Li}D_{Rj}\overline{D}%
_{Rk}D_{Ll}|y_{b}^{(0)}|^{2}V_{3i}^{*}U_{3j}U_{3k}^{*}V_{3l}\frac{2}{%
m_{A}^{2}}  \nonumber \\
&&+\frac{1}{2}\overline{D}_{Li}D_{Rj}\overline{D}%
_{Lk}D_{Rl}|y_{b}^{(0)}|^{2}V_{3i}^{*}U_{3j}V_{3k}^{*}U_{3l}\ \frac{1}{m_{h}^{2}\tan
^{2}\beta }\label{4fer} \\
&&+\frac{1}{2}\overline{D}_{Ri}D_{Lj}\overline{D}%
_{Rk}D_{Ll}|y_{b}^{(0)}|^{2}U_{3i}^{*}V_{3j}U_{3k}^{*}V_{3l}\frac{1}{m_{h}^{2}\tan
^{2}\beta }  \nonumber
\end{eqnarray}

This interaction allows us to calculate the Higgs exchange contribution to all $%
\Delta F=2$ processes in the quark sector. To simplify things further we take $m_{H}^{2}$ and $%
m_{A}^{2}$ in the ballpark of 500 GeV, so that $m_Z^2\ll m_H^2,\,m_A^2\ll m_h^2\tan^{2}\beta$ and the first term in Eq. (\ref{4fer}) is presumably the dominant. We neglect also possible interference of
interaction (\ref{4fer}) with supersymmetric and SM box diagrams. Performing
standard QCD running of $\overline{q}_{L}q_{R}\overline{q}_{R}q_{L}$ operator from
500 GeV down to the scale of 5 GeV and 1 GeV and taking  matrix elements
in vacuum insertion approximation, we arrive at the following formulae
for mass splitting in the neutral B-mesons, K-mesons and $\epsilon _{K}$
parameter:

\begin{eqnarray}
\Delta m_{B} &\simeq&1.5\eta _{B}|y_{b}^{(0)}|^{2}\frac{m_{B}f_{B}^{2}}{%
m_{H}^{2}}|V_{31}^{*}U_{33}V_{33}U_{31}^{*}|,  \label{B} \\
\Delta m_{K} &=&\eta _{K}|y_{b}^{(0)}|^{2}\frac{m_{K}f_{K}^{2}}{m_{H}^{2}}%
\left( \frac{m_{K}}{m_{s}+m_{d}}\right)
^{2}|V_{31}^{*}U_{32}V_{32}U_{31}^{*}|,  \label{K} \\
\epsilon _{K} &=&\eta _{K}|y_{b}^{(0)}|^{2}\frac{m_{K}f_{K}^{2}}{\sqrt{2}%
\Delta m_{K}m_{H}^{2}}\left( \frac{m_{K}}{m_{s}+m_{d}}\right) ^{2}\mbox{Im}%
\left( V_{31}^{*}U_{32}V_{32}U_{31}^{*}\right) .  \label{E}
\end{eqnarray}

Here $\eta _{K}\simeq 5$ and $\eta _{B}\simeq 2.5$ are the QCD renormalization coefficients and the Yukawa coupling $%
y_{b}^{(0)}$ is taken at the scale of 500 GeV (At one-loop accuracy, the operator $(\bar q_L q_R)(\bar q_R q_L)$ does not mix with other possible operators). The decay constant are normalized in such a way that $f_K=160$ MeV and $f_B=200$ MeV. The comparison with
experimental data yields the constraints on the combination of the
off-diagonal elements of $V$ and $U$, Higgs mass and Yukawa coupling $%
y_{b}^{(0)}$. These constraints are summarized in Table 1.

\smallskip

\begin{center}
Table 1.

\vspace{1cm}

\large
$
\begin{tabular}{|c|c|}
\hline
$\Delta m_{B}$ & $\left( \frac{\mbox{{\normalsize 500 GeV}}}{m_{H}}\right)
^{2}|y_{b}^{(0)}|^{2}|V_{31}^{*}U_{33}V_{33}U_{31}^{*}|<1.1\cdot 10^{-7}$ \\ 
\hline
$\Delta m_{K}$ & $\left( \frac{\mbox{{\normalsize 500 GeV}}}{m_{H}}\right)
^{2}|y_{b}^{(0)}|^{2}|V_{31}^{*}U_{32}V_{32}U_{31}^{*}|<1.2\cdot 10^{-9}$ \\ 
\hline
$\epsilon _{K}$ & $\left( \frac{\mbox{{\normalsize 500 GeV}}}{m_{H}}\right)
^{2}|y_{b}^{(0)}|^{2}\mbox{Im}\left( V_{31}^{*}U_{32}V_{32}U_{31}^{*}\right)
<3.5\cdot 10^{-12}$ \\ \hline
\end{tabular}
$
\end{center}
\normalsize

\vspace{1cm}

If $y_{b}^{(0)}\sim y_{t}^{(0)}\simeq 1$ (large $\tan \beta $ regime), it
implies very strong constraints on the off-diagonal elements of $V$ and $U$.
These constraints are 'oblique', i.e. the specifics of the soft-breaking
sector in different supersymmetric models enters only through $%
V_{13},U_{23}^{*}$, etc. They are strongly violated if we take $%
|y_{b}^{(0)}|\simeq 1,$ $|V_{31}|\sim |U_{31}|\sim |V_{ts}|$ and $%
|V_{32}|\sim |U_{32}|\sim |V_{td}|$. It shows that unlike the eigenvalues of
the mass matrices, the mixing angles should not receive more than 10\%
renormalization from the threshold corrections if both $V$ and $U$ matrices
are nontrivial. The numbers on the r.h.s. in Table 1 coincide with the
numbers quoted usually as the limits on the ''superweak interaction'' \cite{Wolf}.
Moreover, the tightest constraint from $\epsilon _{K}$ suggests the
possibility of having nearly real $V_{KM}$ with CP-violation in the Kaon
sector coming from the small phases of the order $10^{-2}$ in $V$ and $U$.

For the generic form of the soft-breaking sector, the limits from Table
1 can be converted into the limits on the off-diagonal entries in the Down
squark mass matrices ${\bf M}_{Q}^{2}$ and ${\bf M}_{D}^{2}$. Treating these
entries as the mass insertions on the scalar quark lines and taking also $(%
{\bf M}_{Q}^{2})_{ii}\sim ({\bf M}_{D}^{2})_{ii}\sim m_{\lambda }^{2}=m^{2}$%
, we are able to calculate $V_{ij}$ and $U_{ij}$, 
\begin{eqnarray}
V_{3i} &=&\frac{1}{3}\frac{\alpha _{s}v}{3\pi (v_{d}+(\mu /m)(\alpha
_{s}/3\pi )v)}\frac{\mu }{m}\frac{({\bf M}_{Q}^{2})_{3i}}{m^{2}}  \label{VU}
\\
U_{3i} &=&\frac{1}{3}\frac{\alpha _{s}v}{3\pi (v_{d}+(\mu /m)(\alpha
_{s}/3\pi )v)}\frac{\mu }{m}\frac{({\bf M}_{D}^{2})_{3i}}{m^{2}}  \nonumber
\end{eqnarray}

Here we keep only gluino exchange contribution and neglect diagrams with
charginos and neutralinos. The combination $v_{d}+(\mu /m)(\alpha _{s}/3\pi
)v$ is the 33 element of the mass matrix $\sqrt{2}M_{d}$, and its moduli
squared corresponds to the observable $2m_{b}^2$ at the scale $m$. This allows
us to modify Eq. (\ref{VU}) so that the combinations of the mixing angles,
relevant for $\Delta B=2$ and $\Delta S=2$ processes, look as follows: 
\begin{eqnarray}
|V_{31}U_{31}| &=&\left| y_{b}^{(0)}\frac{1}{3}\frac{\alpha _{s}}{3\pi y_{SM}%
}\frac{\mu }{m}\right| ^{2}\left| \frac{({\bf M}_{Q}^{2})_{31}}{m^{2}}\frac{(%
{\bf M}_{D}^{2})_{31}}{m^{2}}\right|  \label{gen} \\
V_{31}^{*}U_{32}V_{32}U_{31}^{*} &=&\left| y_{b}^{(0)}\frac{1}{3}\frac{%
\alpha _{s}}{3\pi y_{SM}}\right| ^{4}\left( \frac{\mu }{m}\right) ^{4}\frac{(%
{\bf M}_{Q}^{2})_{31}}{m^{2}}\frac{({\bf M}_{D}^{2})_{31}}{m^{2}}\frac{({\bf %
M}_{Q}^{2})_{32}}{m^{2}}\frac{({\bf M}_{D}^{2})_{32}}{m^{2}}.  \nonumber
\end{eqnarray}
Here $y_{SM}$ is the SM Yukawa coupling of the $b$-quark, $y_{SM}(m)=\sqrt{2}%
m_{b}(m)/v$. It compensates the smallness coming from the high power of the loop
factor, $\alpha _{s}/3\pi $. We summarize the constraints on the $SUSY$
parameter space for the case of the generic form of the soft-breaking in
Table 2.

\smallskip

\begin{center}
Table 2.

\vspace{1cm}

\large
\begin{tabular}{|c|c|}
\hline
$\Delta m_{B}$ & $\left( \frac{\mbox{{\normalsize 500 GeV}}}{m_{H}}\right)
^{2}|y_{b}^{(0)}|^{4}\left| \frac{\mu }{m}\right| ^{2}\left| \frac{({\bf M}%
_{dL}^{2})_{31}}{m^{2}}\frac{({\bf M}_{dR}^{2})_{31}}{m^{2}}\right|
<1.4\cdot 10^{-6}$ \\ \hline
$\Delta m_{K}$ & $\left( \frac{\mbox{{\normalsize 500 GeV}}}{m_{H}}\right)
^{2}|y_{b}^{(0)}|^{6}\left| \frac{\mu }{m}\right| ^{4}\left| \frac{({\bf M}%
_{Q}^{2})_{31}}{m^{2}}\frac{({\bf M}_{D}^{2})_{31}}{m^{2}}\frac{({\bf M}%
_{Q}^{2})_{32}}{m^{2}}\frac{({\bf M}_{D}^{2})_{32}}{m^{2}}\right| <2.0\cdot
10^{-7}$ \\ \hline
$\epsilon _{K}$ & $\left( \frac{\mbox{{\normalsize 500 GeV}}}{m_{H}}\right)
^{2}|y_{b}^{(0)}|^{6}$Im$\left( \left( \frac{\mu }{m}\right) ^{4}\frac{({\bf %
M}_{Q}^{2})_{31}}{m^{2}}\frac{({\bf M}_{D}^{2})_{31}}{m^{2}}\frac{({\bf M}%
_{Q}^{2})_{32}}{m^{2}}\frac{({\bf M}_{D}^{2})_{32}}{m^{2}}\right) <5.5\cdot
10^{-10}$ \\ \hline
\end{tabular}
\end{center}
\normalsize

\vspace{1cm}

We take $m_{\lambda }$ as the real parameter and therefore we have to include
possible phase of $\mu $ in the $\epsilon _{K}$-constraint. These
constraints look somehow relaxed as compared to those from Table 1. The
reason for that is the presence of additional combinatorial $1/3$ in Eqs.
(\ref{VU}) as compared to the $\alpha_s/(3\pi)$ factor, characterizing the renormalization
of eigenvalues. This combinatorial factor arises from the $n=1$ order of
mass insertions in the squark line and leads to one order of magnitude
suppression for $\Delta m_{B}$ and two order of magnitudes for $\Delta m_{K}$
and $\epsilon _{K}$. The fourth and sixth power of $y_{b}^{(0)}$ in Table 2 at
low and intermediate $\tan \beta $ correspond to the $\tan ^{4}\beta $ and $%
\tan ^{6}\beta $ growth of Higgs-mediated FCNC amplitudes claimed earlier in
the Introduction.

The constraints on the soft-breaking terms quoted in Table 2 are very
sensitive to the value of $y_{b}^{(0)}$ and $\mu /m$ ratio. Nevertheless, they
provide valuable limits on the squark flavour sector in the case of
large $\tan \beta $, complementary and sometimes much stronger than those
from the box diagram if $y_{b}^{(0)}\sim \mu /m$ $\sim 1$ \cite{FCNC}. The exception are
the limits on $({\bf M}_{Q}^{2})_{12},$ $({\bf M}_{D}^{2})_{12} $ entries in
the squark mass matrices. The limits on these entries provided by Higgs
exchange are relaxed by some power of the ratio $y_{s}^{(0)}/y_{b}^{(0)}$and
we do not quote them here.

Many specific $SUSY$ models predict certain patterns for the squark mass
matrices so that FCNC amplitudes can be calculated in more details and where the comparison with the box diagram contribution can be made. 

\section{ Constraints on the soft-breaking terms in different SUSY models}

In what follows we consider the cases of Minimal supergravity model, $SUSY$ $%
LR$ and $SUSY$ $SO(10)$ models and so called ''effective supersymmetry''.

{\em 1. Minimal supergravity model}

It is customary to assume, at the scale of the breaking, that the following, very
restrictive conditions are fulfilled: 
\begin{eqnarray}
{\bf M}_{Q}^{2} &=&m_{Q}^{2}{\bf 1};\;\;{\bf M}_{D}^{2}=m_{D}^{2}{\bf 1};\;\;%
{\bf M}_{U}^{2}=m_{U}^{2}{\bf 1}\;\;\mbox{''degeneracy''}\;  \label{eq:deg}
\\
{\bf A}_{u} &=&A_{u}{\bf Y}_{u};\;\;{\bf A}_{d}=A_{d}{\bf Y}_{d}\;\;\;%
\mbox{''proportionality''}.  \label{eq:prop}
\end{eqnarray}
These conditions, if held, ensure that the physics of flavour comes entirely
from the superpotential. We would refer to this possibility as to the
supergravity scenario. Further RG evolution of the soft-breaking parameters,
from the scale of the $SUSY$ breaking down to the electroweak scale, induces
significant off-diagonal terms in ${\bf M}_{Q}^{2}$ whereas ${\bf M}_{D}^{2}$
and ${\bf M}_{U}^{2}$ stay essentially flavour blind. As a result, no
significant right-handed rotation angles can be generated at the threshold,
i.e. $|U_{ij}|\sim \delta _{ij}$. The left-handed squark mass matrix is
nontrivial, though, and, as a result, 13 and 23 elements of the rotation
matrix $V$ are proportional to $V_{td}$ and $V_{ts}$ times the
characteristic splitting in left-handed sector induced by RG running, $%
y_{t}^{2}3\ln (\Lambda /m)/(8\pi ^{2})$. Thus, the Higgs exchange would
induce the operator $\overline{d}_{L}b_{R}\overline{d}_{L}b_{R}$
proportional to $V_{td}^{2}$ which is relevant for the $B$-meson splitting.
Unfortunately, in the case of the Higgs exchange this operator is suppressed
by $\tan ^{-2}\beta $ (See Eq. (\ref{4fer})) and cannot compete with the box
diagram. For the $\Delta S=2$ processes the degree of suppression is even
higher, since the Higgs mediation mechanism involves $\left(
y_{s}^{(0)}/y_{b}^{(0)}\right) ^{2}$.

{\em 2. Effective supersymmetry\newline
\ \ \ }The departure from the strict conditions of degeneracy and
proportionality may occur in several ways which usually implies significant $%
SUSY$ contributions to FCNC amplitudes. Here we would skip the theoretical
justification for certain choices of the soft-breaking parameters going
directly to the phenomenological consequences related with Higgs-mediated
FCNC. 

We turn now to the ''effective supersymmetry'' picture \cite{ES1} which has
much less degrees of freedom and where flavour physics can be formulated in
a more definitive way. In the squark mass matrices, diagonalized by unitary
transformations $U$ and $V$, 
\begin{equation}
{\bf M}_{Q}^{2}=\widetilde{V}^{\dagger }\left( 
\begin{array}{ccc}
m_{1}^{2} & 0 & 0 \\ 
0 & m_{2}^{2} & 0 \\ 
0 & 0 & m_{3}^{2}
\end{array}
\right) \widetilde{V};\;\;{\bf M}_{D}^{2}=\widetilde{U}^{\dagger }\left( 
\begin{array}{ccc}
m^{\prime }{}_{1}^{2} & 0 & 0 \\ 
0 & m^{\prime }{}_{2}^{2} & 0 \\ 
0 & 0 & m^{\prime }{}_{3}^{2}
\end{array}
\right) \widetilde{U},  \label{ES}
\end{equation}
the eigenvalues $m_{1}^{2}$, $m_{2}^{2}$, $m^{\prime }{}_{1}^{2}$, $%
m^{\prime }{}_{2}^{2}$ are taken to be in the multi-TeV scale and eventually
decoupled from the rest of particles. At the same time, the squarks from the
third generation are believed to be not heavier than $1$ TeV and weakly
coupled to the first and second generations of quarks to avoid the excessive
fine-tuning in the radiative corrections to the Higgs potential and suppress
FCNC contribution to the Kaon mixing. The advantage of this approach in our
case is that the same loop integral stands for the renormalization of the
Yukawa couplings and mixing angles. Moreover, it is easy to see that
matrices $\widetilde{V}$ and $\widetilde{U}$ in Eq. (\ref{ES}) and $V$ and $U
$ in Eq. (\ref{Md}) are closely related: 
\begin{eqnarray}
\nonumber|V_{31}U_{31}| =\left| y_{b}^{(0)}\frac{\alpha _{s}}{3\pi y_{SM}}\frac{\mu 
}{m_{3}}F(m_{\lambda }/m_{3})\right| ^{2}\left| \widetilde{V}_{31}\widetilde{%
U}_{31}\right| \simeq 0.7\left( \frac{y_{b}^{(0)}\mu }{m_{3}}\right)
^{2}\left| \widetilde{V}_{31}\widetilde{U}_{31}\right|  \\
V_{31}^{*}U_{32}V_{32}U_{31}^{*} =\left| y_{b}^{(0)}\frac{\alpha _{s}}{%
3\pi y_{SM}}F(m_{\lambda }/m_{3})\right| ^{4}\left( \frac{\mu }{m_{3}}%
\right) ^{4}\widetilde{V}_{31}^{*}\widetilde{U}_{32}\widetilde{V}_{32}%
\widetilde{U}_{31}^{*}\simeq 0.5\left( \frac{y_{b}^{(0)}\mu }{m_{3}}\right)
^{4}\widetilde{V}_{31}^{*}\widetilde{U}_{32}\widetilde{V}_{32}\widetilde{U}%
_{31}^{*},  \nonumber \end{eqnarray}
\begin{equation} 
F(x) =\frac{2x}{1-x^{2}}+\frac{2x^{3}\ln x{^{2}}}{(1-x^{2})^{2}}, 
\label{F}
\end{equation}
and $F(1) =1$ is substituted in the first and second lines of Eq. (\ref{ES}).
These formulae allows us to limit the mixing angles in $\widetilde{V}$ and $%
\widetilde{U}$ matrices; for $F\simeq 1$ and $y_{b}^{(0)}\simeq 1$ they are
almost as strong as the constraint on $V$ and $U$ from Table 1.
Numerically, it is important that the combinatorial suppression from Eq. (%
\ref{gen}), $\left( 1/3\right) ^{2}$ and $\left( 1/3\right) ^{4}$, does not
apply here. Table 3 summarizes the limits on the parameter space of the
model, provided by $\Delta F=2$ amplitudes mediated by Higgses

\smallskip
\vspace{1cm}

\begin{center}
Table 3

\vspace{1cm}

\large
\begin{tabular}{|c|c|}
\hline
$\Delta m_{B}$ & $\left( \frac{\mbox{{\normalsize 500 GeV}}}{m_{H}}\right)
^{2}|y_{b}^{(0)}|^{4}\left| \frac{\mu }{m_3}\right| ^{2}\left| \widetilde{V}%
_{31}\widetilde{U}_{31}\right| <1.6\cdot 10^{-7}$ \\ \hline
$\Delta m_{K}$ & $\left( \frac{\mbox{{\normalsize 500 GeV}}}{m_{H}}\right)
^{2}|y_{b}^{(0)}|^{6}\left| \frac{\mu }{m_3}\right| ^{4}\left| \widetilde{V}%
_{31}^{*}\widetilde{U}_{32}\widetilde{V}_{32}\widetilde{U}_{31}^{*}\right|
<2.4\cdot 10^{-9}$ \\ \hline
$\epsilon _{K}$ & $\left( \frac{\mbox{{\normalsize 500 GeV}}}{m_{H}}\right)
^{2}|y_{b}^{(0)}|^{6}$Im$\left( \left( \frac{\mu }{m_3}\right) ^{4}\widetilde{V%
}_{31}^{*}\widetilde{U}_{32}\widetilde{V}_{32}\widetilde{U}_{31}^{*}\right)
<7.0\cdot 10^{-12}$ \\ \hline
\end{tabular}
\normalsize
\smallskip
\end{center}

\vspace{1cm}

These constraints are taken at $F\simeq 1$. In Fig. 2 we plot $F$ and $F^{4}$
as a function of the ratio $m_{\lambda }/m_{3}$ . It is interesting to note
that Higgs-mediated amplitudes can probe the region of the parameter space
where gluino is considerably heavier than stop and sbottom. All other
observables related with dipole amplitudes drop off very fast with $%
m_{\lambda }$ and do not produce significant constraints in this part of the parameter space. The comparison of the box diagram contribution to the $B-%
\overline{B}$ mixing with the Higgs-mediated mechanism provides the value of
the critical $\left( \tan \beta \right) _{cr}$ above which the Higgs
exchange dominates: 
\begin{equation}
\left( \tan \beta \right) _{cr}=17
\label{tes}
\end{equation}

The numbers from Table 3 suggest also that the observable value of $\epsilon_K$ parameter in this model can be achieved through the small imaginary phases of $\mu$ and/or $\widetilde{V}_{ij}$, $\widetilde{U}_{ij}$. 

{\em 3. Supersymmetric SO(10) and left-right models\newline
\ \ \ } Another example of $SUSY$ theory that we would like to consider here
are the supersymmetric $SO(10)$ \cite{so10,BHS} and the left-right models \cite{LR}. In  
$SO(10)$ model, left- and right-handed quark superfields are unified at some scale $%
\Lambda _{GUT}$
into the same multiplet. Therefore, above the scale of the unification the RG evolution of ${\bf M}_{Q}^{2}$ and ${\bf M}_{D}^{2}$
should be the same and these matrices will be equally affected by large Yukawa
couplings of the third generation. Therefore, if ${\bf M}_{Q}^{2}$ exibits
non-trivial flavour structure, so does ${\bf M}_{D}^{2}$. As a result, the
squark degeneracy condition is violated both in the left- and right-handed
sector and the characteristic splitting is proportional to $y_{t}^{2}\ln (\Lambda _{Plank}^{2}/\Lambda _{GUT}^{2})$.  In the basis where $%
{\bf Y}_{d}^{(0)}$ matrix is diagonal, the off-diagonal entries in the
squark mass matrices can be related to the mixing angles of the rotation
matrices between two basis given by diagonal ${\bf Y}_{d}^{(0)}$ and diagonal 
${\bf Y}_{u}^{(0)}$. Further RG evolution, from $\Lambda _{GUT}$ to the weak
scale, deviates ${\bf M}_{Q}^{2}$ from ${\bf M}_{D}^{2}$ but the nontrivial
part of ${\bf M}_{D}^{2}$ survives and generates very important
phenomenological consequences. It is easy to see, though, that somehow
milder assumption about left-right symmetry in the theory with $%
SU(2)_{L}\times \ SU(2)_{R}\times U(1)$ gauge group can be responsible for
exactly the same phenomenology \cite{Posp}. The role of $\Lambda _{GUT}$ in this case is
played by $\Lambda _{LR}$, the scale where left-right symmetry gets
spontaneously broken and heavy right-handed gauge bosons decouple. 

To study the Higgs-mediated FCNC amplitudes we need to know the initial form
of the Yukawa matrix  ${\bf Y}_{d}^{(0)}$ in the basis where ${\bf Y}%
_{u}^{(0)}$ is diagonal. Two choices seem to be justified: hermitian ${\bf Y}%
_{d}^{(0)}$ \cite{MR} or complex symmetric ${\bf Y}_{d}^{(0)}$ \cite{DH}.
Depending on this choice, right-handed mixing matrix is either exactly the
same as KM matrix or equal to the transposed KM matrix times the diagonal
matrix with two new physical phases in it.
Now we are ready to
calculate relevant combinations of mixing angles in terms of the KM matrix elements $V_{td}$ and $V_{ts}$:
\begin{eqnarray}
|V_{31}U_{31}| &=&\left| y_{b}^{(0)}\frac{\alpha _{s}}{3\pi y_{SM}}\frac{\mu 
}{m_{3}}G(m_{\lambda },m_{3},m)\right| ^{2}\left| V_{td}\right| ^{2}\simeq
7\cdot 10^{-5}\left( \frac{y_{b}^{(0)}\mu }{m_{3}}G(m_{\lambda
},m_{3},m)\right) ^{2}  \nonumber \\ \label{LR1}
\left| V_{31}^{*}U_{32}V_{32}U_{31}^{*}\right|  &=&\left| y_{b}^{(0)}\frac{%
\alpha _{s}}{3\pi y_{SM}}\frac{\mu }{m_{3}}G(m_{\lambda },m_{3},m)\right|
^{4}\left| V_{td}\right| ^{2}\left| V_{ts}\right| ^{2} \\
&\simeq &8\cdot 10^{-8}\left( \frac{y_{b}^{(0)}\mu }{m_{3}}G(m_{\lambda
},m_{3},m)\right) ^{4},  \nonumber
\end{eqnarray}
where we take $V_{td}\simeq 0.01$ and $V_{ts}\simeq 0.04$. 
The invariant function $G(m_{\lambda },m_{3},m)$ still depends on many parameters: rihgt-handed sbottom mass, left-handed sbottom mass, masses of squarks from first and second generations and gluino mass. Substituting (\ref{LR1}) into the general limits from  Table 1, we obtain the following constraints on the parameter space of the $SO(10)$ and left-right symmetric types of models:

\begin{center}
Table 4

\vspace{1cm}

\large
\begin{tabular}{|c|c|}
\hline
$\Delta m_{B}$ & $\left( \frac{\mbox{{\normalsize 500 GeV}}}{m_{H}}\right)
^{2}|y_{b}^{(0)}|^{4}\left( \frac{\mu }{m_3}G\right) ^{2} <1.6\cdot 10^{-3}$ \\ \hline
$\Delta m_{K}$ & $\left( \frac{\mbox{{\normalsize 500 GeV}}}{m_{H}}\right)
^{2}|y_{b}^{(0)}|^{6}\left( \frac{\mu }{m_3}G\right) ^{4}
<1.5\cdot 10^{-2}$ \\ \hline
$\epsilon _{K}$ & $\left( \frac{\mbox{{\normalsize 500 GeV}}}{m_{H}}\right)
^{2}|y_{b}^{(0)}|^{6}\left( \frac{\mu }{m_3}G\right) ^{4} \sin(\phi_s-\phi_d)
<4.4\cdot 10^{-4}$ \\ \hline
\end{tabular}
\normalsize
\smallskip
\end{center}

\vspace{1cm}

The last row in Table 4 refers to the case of the complex symmetric mass matrices and $\phi_s$, $\phi_d$ are the new physical phases associated with the right-handed mixing matrix. The dependence of these phases is the same as in the case of the box-diagram \cite{BHS}. The significance of these constraints depends very strongly on $G$. To illustrate this we take the masses of the first and second generation of squarks to be equal to gluino mass and take also the same mass $m_3$ for the left- and right-handed sbottom. Then, the function $G$ can be reexpressed in terms of the function F, introduced earlier in Eq. (\ref{F}),
\begin{equation}
G(m_{\lambda},m_{3},m) = F(m_{\lambda}/m_3)-\frac{m_3}{m}
F(m_{\lambda}/m)=F(m/m_3)-\frac{m_3}{m}
\end{equation}
In Fig. 3  we plot $G$ and $G^4$ as a function of the ratio $m/m_3$. As in the previous case, there is a significant sensitivity to the part of the parameter space where gluino and first generation of squarks are significantly heavier than $m_3$. The comparison with the box diagram is also very much dependent of the  $m/m_3$ ratio. If this ratio is large $m/m_3>2$, the critical value of $tan\beta$ is approximately the same as in the case of completely decoupled first and second generations of squarks (\ref{tes}). 

\section{Conclusions}

We have considered in details FCNC processes mediated by Higgs particles in
the supersymmetric models with large $\tan \beta$ . 
Large $\tan \beta $ and $\mu \sim m_{sq}\sim m_{\lambda }$ invoke
significant renormalization of $M_{d}$, mass matrix for the Down-type of
quarks. It is evident that the mixing angles can also acquire
additional contributions from the threshold corrections if flavour can be changed in the squark sector. When this renormalization occurs both in left- and right- handed rotation matrices, it
generates flavour changing operators $\overline{d}_{L}b_{R}\overline{d}%
_{R}b_{L}$ and $\overline{d}_{L}s_{R}\overline{d}_{R}s_{L}$ mediated by
heavy Higgs particles $H$ and $A$. The coefficients in front of these
operators are further enhanced by RG evolution to the low-energy scale and,
in the case of neutral kaons, by chiral enhancement factor in the matrix
elements. 

For the models with $\tan \beta \sim O(1)$ the Higgs-mediated FCNC amplitudes are truly marginal because they appear at two or moore loops. The $SUSY$ contribution to FCNC in this case is given
by box diagrams. However, Higgs-mediated FCNC amplitudes strongly depend 
on $\tan \beta $; they grow as the fourth
and sixth power of $\tan\beta$ for $\Delta B=2$ and $\Delta S=2$ processes respectively. As a result, Higgs-mediated amplitudes can match $SUSY$ box diagrams when $\tan \beta \sim 20$ and   
become the dominant mechanism for FCNC when $\tan \beta $ $\sim m_{t}/m_{b}$. In latter case, Higgs-mediated FCNC
amplitudes provide significant constraints on the off-diagonal elements in the squark mass
matrices. Very strong limits on the soft-breaking sector are 
formulated here in Tables 1 and 2. These limits are  complementary to those provided by box diagrams \cite{FCNC}. The constraints from Table 1  are oblique, i.e.
all the details about the soft-breaking sector are confined in the rotation
angles between the mass basis provided by superpotential and the physical
mass basis which includes $v_{u}$-dependent radiative corrections. 

The significance of this mechanism varies from one model to another and depends
mainly on the nontrivial structure of the right-handed squark mass matrix.
For the minimal supergravity scenario this matrix is trivial and Higgs
exchange mechanism is not important for any value of $\tan\beta$. Significant contributions to $\Delta
m_{K}$, $\Delta m_{B}$ and $\epsilon _{K}$ can be induced in other types of $SUSY$ models where right-handed squark mass matrices are similar to the left-handed ones. 

\vspace{0.5cm}

\mbox{\hspace{-0.5cm}}{\large {\bf Acknowledgements}} \newline This work is supported in part by N.S.E.R.C. of Canada.

\newpage
\large{\bf Figure captions }\normalsize

\vspace{1cm}

Figure 1.  The diagrams generating $v_u$-dependent threshold corrections to ${\bf M_d}$.

\vspace{1cm}

Figure 2. In the case of decoupled first and second generation of squarks, the invariant functions $F$ and $F^4$ are plotted against $m_\lambda/m_3$ ratio. The Higgs-mediated FCNC amplitudes are sensitive to large $m_\lambda$.

\vspace{1cm}

Figure 3. $SO(10)$ and left-right $SUSY$ models. $G$ and $G^4$ are plotted as a function of $m/m_3$ ratio taking $m= m_\lambda=m_1=m_2$. 

\vspace{1cm}
\begin{figure}[hbtp]
\begin{center}
\mbox{\epsfxsize=144mm\epsffile{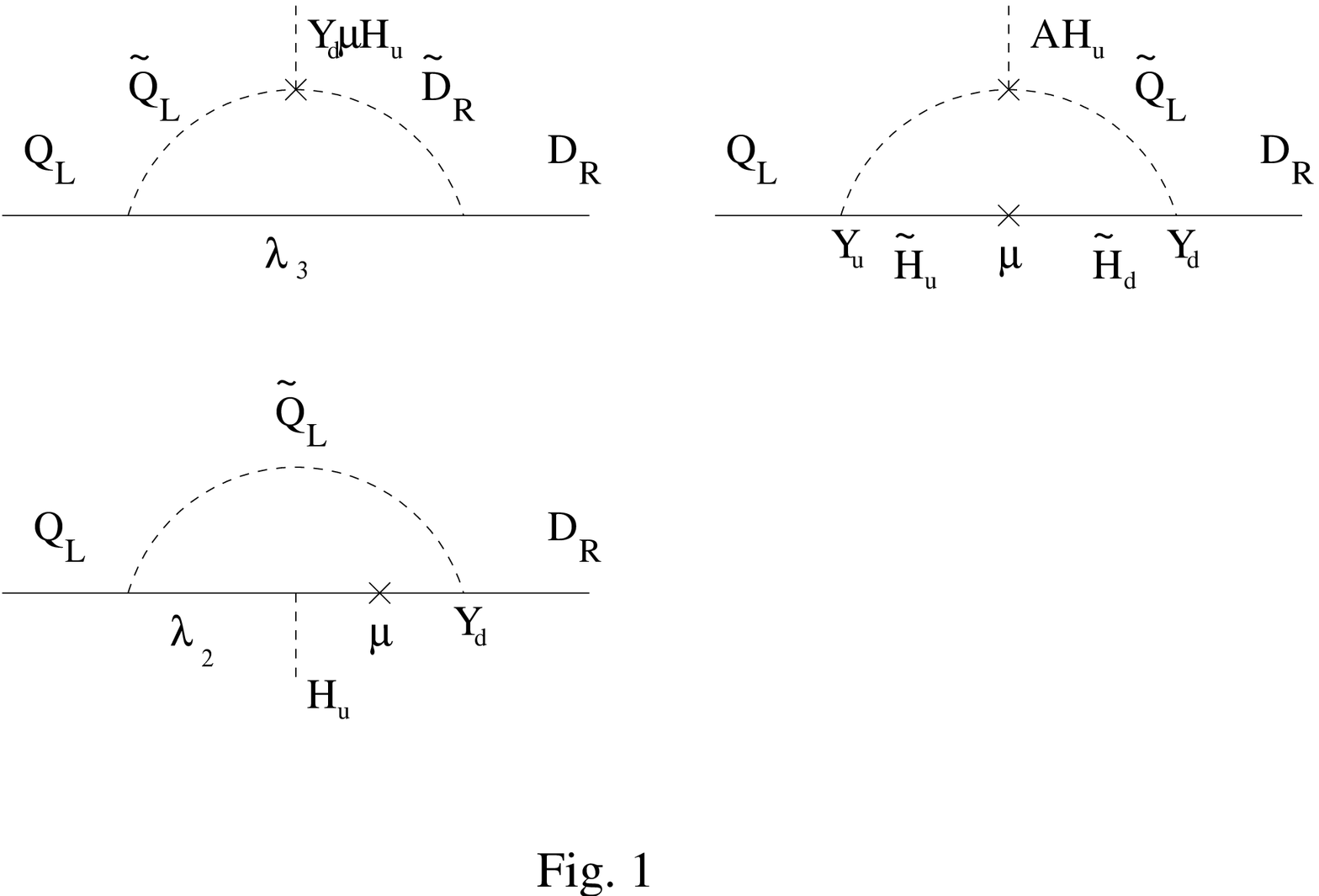}}
\end{center}
\end{figure}

\newpage
\begin{figure}[hbtp]

\mbox{\epsfxsize=80mm\epsffile{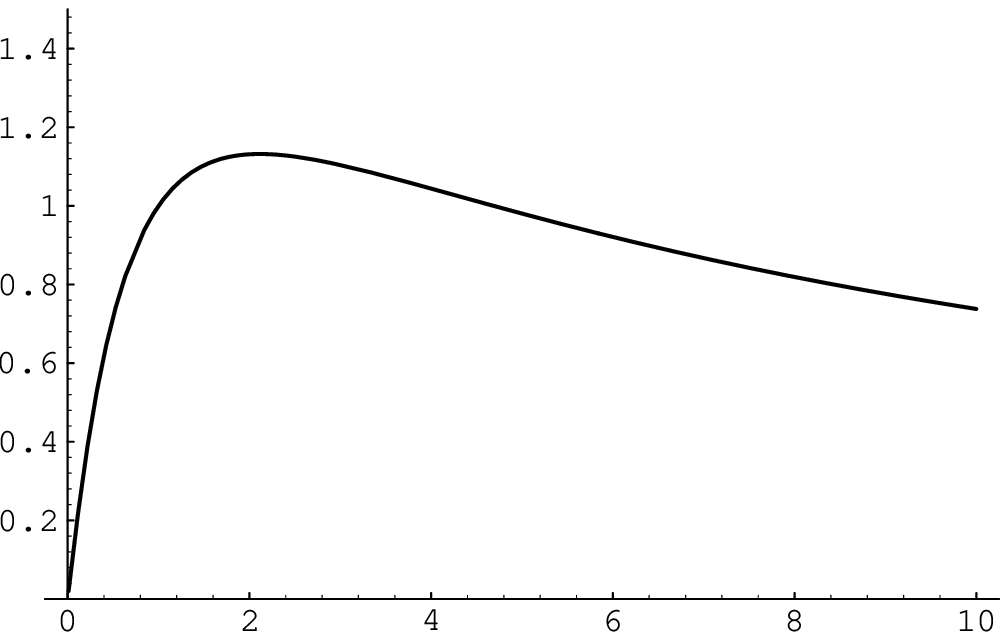}}
\mbox{\epsfxsize=80mm\epsffile{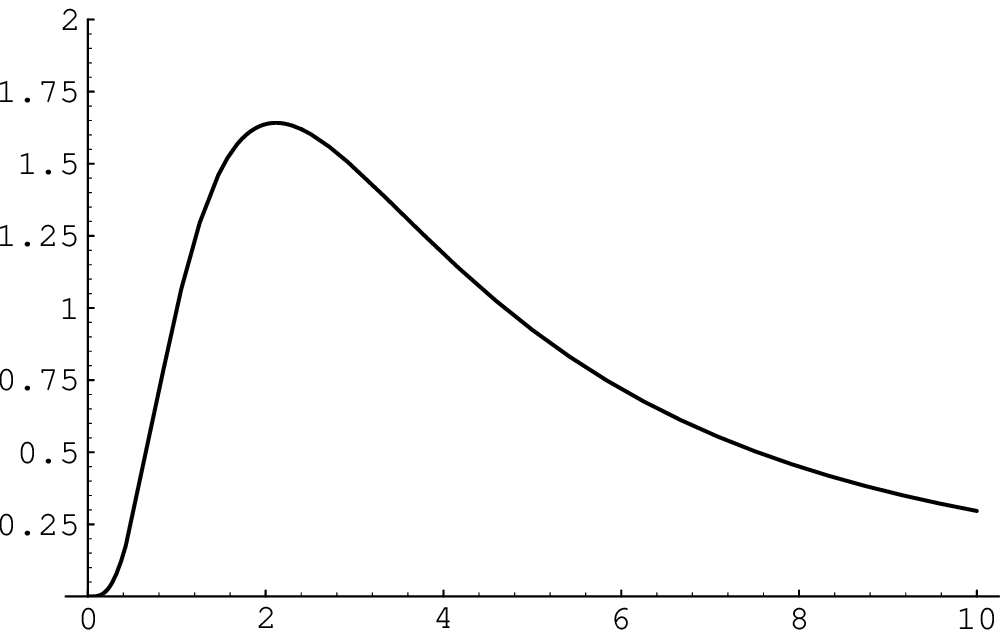}}

\end{figure}

\begin{center}{\large Figure 2}
\end{center}

\begin{figure}[hbtp]

\mbox{\epsfxsize=80mm\epsffile{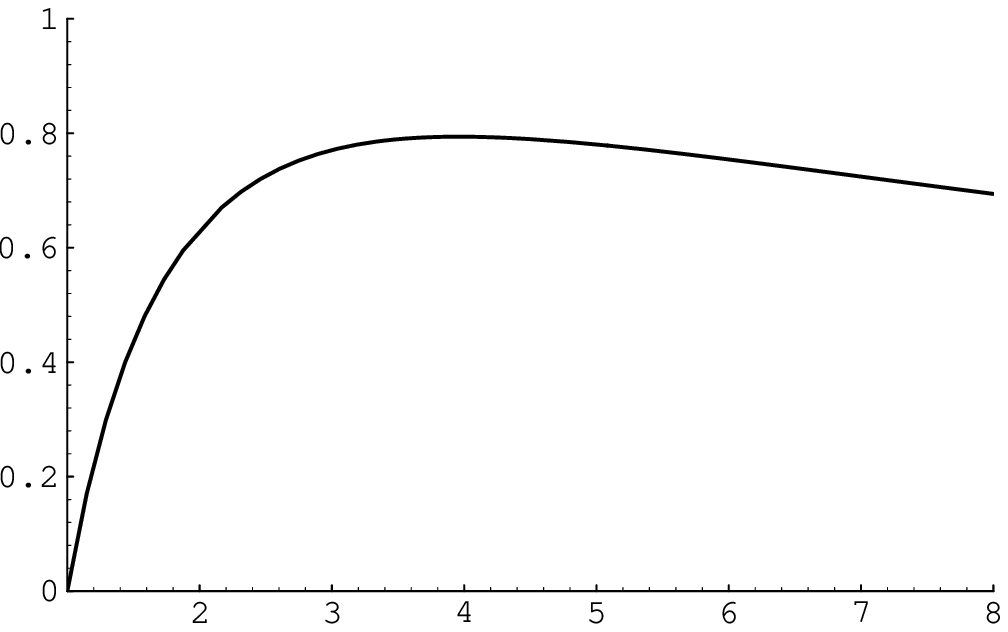}}
\mbox{\epsfxsize=80mm\epsffile{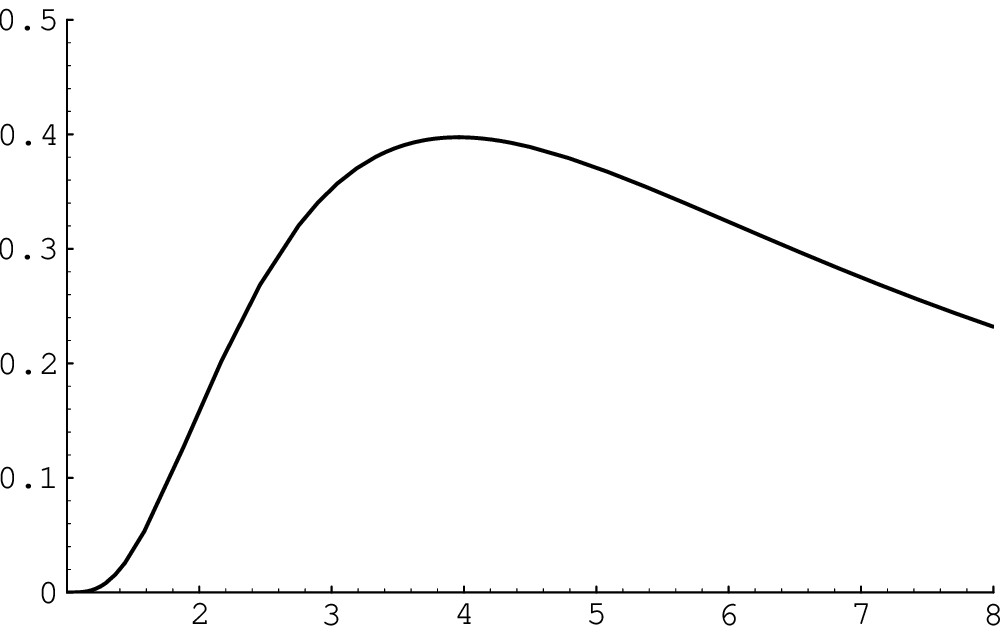}}

\end{figure}
\begin{center}\large Figure 3\end{center}

\end{document}